\documentclass[prl,twocolumn,showpacs,preprintnumbers,amsmath,amssymb]{revtex4}

\begin{document}

\title{A simple thermodynamical witness showing universality of macroscopic entanglement}


\author{Vlatko Vedral}
\email{v.vedral@quantuminfo.org}
\affiliation{The School of Physics and Astronomy, University of Leeds,
Leeds LS2 9JT, England and \\
Centre for Quantum Technologies, National University of Singapore, 3 Science Drive 2, Singapore 117543\\
Department of Physics, National University of Singapore, 2 Science Drive 3, Singapore 117542}

\begin{abstract}
We show that if the ground state entanglement exceeds the total entropy of a given system, then this system is in an entangled state. This is a universal entanglement witness that applies to any physical system and yields a temperature below which we are certain to find some entanglement. Our witness is then applied to generic bosonic and fermionic many body systems to derive the corresponding "critical" temperatures that have a very broad validity.
 \bigskip
\end{abstract}
\maketitle



Entanglement has recently been extensively investigated and even detected in various many-body systems \cite{Amico}. Here we would like to make a claim that all these different results in fact conform to a certain universal behaviour that can be uncovered using very simple thermodynamical arguments. In order to do so, we will have to choose one particular way of thinking about entanglement, that has clear connections with thermodynamics. We will look at the trade-off between the amount of entanglement in the ground state of a given system, as quantified by the relative entropy of entanglement, and the mixedness of the system at a certain temperature, as quantified by its total entropy.

Suppose that we are given a thermal state $\rho_T = p|\Psi_0\rangle\langle \Psi_0 | + (1-p) \rho_{rest}$, where $|\Psi_0\rangle$ is the ground state, $p=\exp(-E_0/kT)/Z$ is the usual Boltzmann weight and $\rho_{rest}$ involves all higher levels. A very simple entanglement witness can now be derived by noting that if
\begin{equation}
S(|\Psi_0\rangle ||\rho_T) < S(|\Psi_0\rangle ||\rho_{sep})=E(|\Psi_0\rangle)
\end{equation}
where $S(\sigma ||\rho)$ is the quantum relative entropy \cite{Umegaki}, then the state $\rho_T$ must be entangled (as it is closer to $|\Psi_0\rangle$ than the closest separable state, which we denoted as $\rho_{sep}$). $E(\rho)$ is the relative entropy of entanglement of $\rho$ \cite{Vedral1,Vedral2}. After a few simple steps, the above inequality leads to another inequality, satisfied by entangled thermal states $\rho_T$,
\begin{equation}
-\ln p < E(|\Psi_0\rangle)
\end{equation}
which was used in \cite{Markham} to investigate entanglement of some many-body systems. Exploiting the fact that
(see for example \cite{Landau})
\begin{equation}
p=\frac{e^{-E_0/kT}}{Z} = e^{-(E_0+kT\ln Z)/kT} \geq e^{-(U+F)/kT}=e^{-S/k}\; ,
\end{equation}
where $F=-kT\ln Z$ is the free energy and $S=(F+U)/T$ is the entropy, we finally obtain the inequality
\begin{equation}
S(\rho_T) < k E(|\Psi_0\rangle)
\end{equation}
implying that $\rho_T$ is entangled. We now have a very simple criterion which can be expressed as follows: if the entropy of a thermal state is lower than the relative entropy of its ground state (multiplied by the Boltzmann constant $k$), then this thermal state contains some form of entanglement. We can also adopt the interpretation of relative entropy due to Donald \cite{Donald}. According to this, the relative entropy $S(\sigma ||\rho_T)$ is equal to the free energy gain when we move from the equilibrium state $\rho_T$ to another state $\sigma$. All our inequality then says is this: if moving from the closest separable state to a pure entangled state requires more free energy than moving from a thermal state to the same pure state, then this thermal state must be entangled.

Here we are not really concerned with the type of entanglement we have (e.g. bi-partite or multipartite, distillable or bound), but we only what to confirm that the state is not fully separable. It is also very clear that if the ground state is not entangled, this witness will never detect any entanglement (since entropy is always a non-negative quantity), even though the state may in reality be entangled for some range of temperatures.

The entanglement witness based on entropy, though at first sight very simple, is nevertheless rather powerful as it allows us to talk very generally about temperatures below which we should start to detect entanglement in a very generic solid state system. The behaviour of any system can be derived from its Hamiltonian that specifies all interactions between subsystems. No matter how complicated this Hamiltonian may be, we can always diagonalise it to the simple form $H = \sum_{i=1}^M \omega_i d^{\dagger}_id_i$,
where $\omega_i$s are its $M$ eigen-energies and $d_i,d^{\dagger}_i$ are the annihilation and creation operators for the $i$-th eigen-mode. We will keep the discussion completely general by considering both fermionic and bosonic commutation relations on $d_i$s, as well as completely distinguishable particles (see for example \cite{Fetter}). The free energy is now easily computed to be: $F =  \pm kT \prod_i \ln (1 \mp e^{\beta(\mu -\omega_i)})$,
where the convention will always be that the upper (lower) sign corresponds to bosons (fermions) and $\mu$ is the chemical potential. Entropy then simply follows via the formula: $S=-\partial F/\partial T$, and is equal to
\begin{equation}
S = - \sum_i n_i \ln n_i \mp (1\pm n_i)\ln (1\pm n_i)
\end{equation}
where $n_i = 1/(\exp \beta(\omega_i - \mu) \pm 1)$.
What matters now is the scaling of entropy with $M$ (the number of modes) and $N$ (the average number of particles). This scaling, in turn, depends on the spectrum of the system $\omega_i$ as well as the temperature and the particle statistics. Since entropy is lower at low temperatures, this is the regime where we expect the witness to show entanglement. Let us look at the typical examples of ideal bosonic and fermionic gasses. Non-ideal systems behave very similarly, with some for us unimportant corrections. At low $T$, the entropy scales as (see e.g. \cite{Rubia})
\begin{equation}
S \sim N (\frac{kT}{\tilde \omega_{F,B}})^{p_{F,B}}
\end{equation}
where $F,B$ refer to fermions and bosons respectively, $N$ is the (average) number of particles, $\tilde \omega$ is some characteristic frequency which is a function of the spectrum (its form depends on the details of the system and one particular example will be presented below) and $p\geq 1$. The fact that this form is the same for more general systems is due to what is known as the third law of thermodynamics (see \cite{Landsberg} for example) stating that the entropy has to go to zero with temperature. We now consider how entanglement scales in the ground state for fermions and bosons \cite{Hayashi}. If the number of particles is comparable to the number of modes, this typically means that $E\sim N$. The entropy witness then yields a very simple temperature below which entanglement exists for both fermions and bosons,
\begin{equation}
kT <  \tilde \omega_{F,B}
\end{equation}
This kind of temperature has been obtained in a multitude of different systems, ranging from spin chains, via harmonic chains and to (continuous) quantum fields. Its universality is now justified from a very simple behaviour of entropy at low temperatures.

In the case of higher temperature, $T\geq \tilde \omega$, both bosonic and fermionic systems approximately obey the Maxwell-Boltzmann statistics  of distinguishable systems (in other words, the thermal de Broglie wavelength of each particle is much smaller than the volume it occupies on average). We then obtain the following inequality:
$N\ln T + N(1-\ln \tilde \omega) < E \sim N$,
where $\ln \tilde \omega = \sum \ln \omega_i/N$ (this gives us an intuition of how typical temperatures compare to the spectral frequencies). It is clear that this inequality will never be satisfied, for the range of temperatures we are considering. Entanglement is thus not expected in an ideal classical gas!

We close with two remarks. The first one is that there is nothing special about using the relative entropy in equation (1). We could have used any other distance measure. The second remark is that similar methods can be used to probe entanglement in quantum phase transitions \cite{Sachdev}. These occur at low temperatures (strictly at $T=0$) and are driven, not by temperature, but by other external parameters, such as a uniform magnetic field. It is impossible now to use the entropy of the state as a witness, since this quantity is always zero at zero temperature. We can use instead the fact that if the energy of a given state $|\Psi\rangle$ exceeds in its absolute value the highest expected value for any separable state, then the state $|\Psi\rangle$ must itself be entangled. This method has been exploited in many papers (see, for instance, \cite{Brukner,Toth,Anders}). Usually, however, appearance of entanglement in the ground state of interacting systems is not surprising and is more common than not. Entanglement is also much easier to detect and quantify since we deal with pure states only. Consequently rigorous results on scaling of entanglement at $T=0$ and our entropy witness could then be applied to tell us at which temperatures we expect the relationship between entropy and area  \cite{Cramer} to fail.

The main motivation of this work was to show that there is a universal finite temperature for many-body systems below which entanglement is guaranteed to exist. We should no longer be surprised by the ubiquity of entanglement
in the macroscopic domain.

\textit{Acknowledgments}: The author acknowledges financial support from the Engineering and Physical Sciences Research Council, the Royal Society and the Wolfson Trust in UK as well as the National Research Foundation and Ministry of Education, in Singapore.

\end{document}